\documentclass[aps,twocolumn,showpacs,nofootinbib,superscriptaddress]{revtex4}

\usepackage{graphicx}
\usepackage{amsmath}
\usepackage{amsfonts}
\usepackage{amssymb}
\usepackage{hyperref}

\newcommand{\beq}{\begin{equation}}
\newcommand{\eeq}{\end{equation}}
\newcommand{\bqa}{\begin{eqnarray}}
\newcommand{\eqa}{\end{eqnarray}}

\begin{document}

\title{Economical ontological models for discrete quantum systems}

\author{Ernesto F. Galv\~{a}o}
\affiliation{Instituto de F\'{i}sica, Universidade Federal Fluminense, \\Av. Gal. Milton Tavares de Souza s/n, Niter\'{o}i, 24210-346, RJ, Brazil}

\date{\today}
\begin{abstract}
I use the recently proposed framework of ontological models [Harrigan \textit{et al.}, arXiv:0709.1149v2] to obtain economical models for results of tomographically complete sets of measurements on finite-dimensional quantum systems. I describe a procedure that simplifies the models by decreasing the number of necessary ontic states, and present an explicit model with just 33 ontic states for a qutrit.
\end{abstract}
\date{\today}
\pacs{03.65.Ta, 03.67.-a, 03.67.Ac }
\maketitle

\section{Introduction}

The inherent complexity in the description of quantum states and operations must be studied if we are to understand the foundations of the theory. This study can shed light on the use of quantum systems for information processing, and clarify issues of simulability and compatibility with other theories, such as relativity.

In this context, there have been some recent proposals to describe quantum theory using general probabilistic theories. The operational ``r-p" framework \cite{Hardy01, Mana04, Barrett07, BarnumBLW07} allows us to study general probabilistic theories, of which quantum mechanics is a particular instance. With the interest in quantum information and in simulating complex quantum phenomena, these recent approaches have yielded interesting results bearing on the efficiency with which we can simulate certain classes of quantum operations \cite{Aaronson05, Montina08}. They have also contributed to the understanding of negativity and contextuality in quasi-probability distributions \cite{Spekkens08}.

Harrigan \textit{et al.} \cite{HarriganRA07} have recently proposed the ontological model formalism to describe any compilation of experimental outcomes. This framework is based on realism, and involves only positive probability distributions and indicator functions. I review their formalism in section \ref{sec ontmod}. They presented three types of general models that describe any set of experimental outcomes. In their paper, however, the models are applied only to a few quantum examples that involved discrete sets of preparations (states) and measurements. A similar approach was recently used to obtain asymptotically efficient models for general quantum measurements \cite{DakicSPB08}.

In section \ref{sec ddim} I apply Harrigan \textit{et al.}'s framework to obtain a model for data arising from a tomographically complete set of measurements on any state of a $d$-dimensional quantum system. I describe how one may decrease the complexity of the model (as measured by the number of necessary ontic states), and obtain a model for a qutrit ($d=3$) with just 33 ontic states. Finding such economical models is important for problems such as the study of hidden-variable theories for quantum mechanics, and the search for efficient descriptions of general experimental data (see \cite{HarriganRA07} for further references).

\section{The ontological model formalism}\label{sec ontmod}

In this section I review the ontological model formalism, introduced by Harrigan, Rudolph and Aaronson in \cite{HarriganRA07}. The aim of the formalism is to obtain a model that describes experimental data compiled in a data table. The model uses positive, normalized probability distributions over a certain number of variables, and indicator functions to describe measurements.

A data table is a compilation of the (probabilistic) results of a series of experiments. An experiment consists of a set of $s$ preparation procedures $\mathcal{P}^{(i)}$, where $i=1,2,...,s$, together with a set of $m$ different measurement procedures $\mathcal{M}^{(i)}$, where $i=1,2,...,m$. We can choose the number of outcomes for each measurement to be $d$ (i.e. the same for all measurements), if necessary by padding with null outcomes.

For concreteness, let us see how to describe a data table arising from projective measurements on a qubit. Each preparation represents a quantum state. The measurements we will consider are the set $\{\mathcal{M}^{(1)}=X, \mathcal{M}^{(2)}=Y, \mathcal{M}^{(3)}=Z\}$ (the Pauli matrices). This set of measurements is tomographically complete, meaning that the probabilities associated with these three operators are sufficient to completely characterize the preparations, and hence sufficient also to calculate the probabilities associated with outcomes of any other measurement.

After many measurements of observables $\mathcal{M}^{(j)}$, the probability (relative frequency) of each outcome can be tabulated in a data table. Below we represent three data tables, corresponding to each of the three measurements $\mathcal{M}^{(j)}$. For convenience, the three data tables are represented on top of each other, forming a single, composite data table:
\begin{equation}
\begin{array}{r@{\extracolsep{\fill}}l}
&\begin{array}{cccc}
& \:\:\:\mathcal{P}^{(1)}& \mathcal{P}^{(2)} &   \\
\end{array}\\
\begin{array}{c}
\mathcal{M}^{(1)}_0 = \left|+\right\rangle\left\langle +\right|\\
\mathcal{M}^{(1)}_1=\left|-\right\rangle\left\langle -\right|\\
\mathcal{M}^{(2)}_0=\left|+i\right\rangle\left\langle +i\right|\\
\mathcal{M}^{(2)}_1=\left|-i\right\rangle\left\langle -i\right|\\
\mathcal{M}^{(3)}_0=\left|0\right\rangle\left\langle 0\right|\\
\mathcal{M}^{(3)}_1=\left|1\right\rangle\left\langle 1\right|\\
\end{array}
&\left[\begin{array}{cccc}
 0.93 \:\:& 0.81 \:\:& ... \:\:\\
 0.07 \:\:& 0.19 \:\:& ... \:\:\\\hline
 0.73 \:\:& 0.16 \:\:& ... \:\:\\
 0.27 \:\:& 0.84 \:\:& ... \:\:\\\hline
 0.59 \:\:& 0.30 \:\:& ... \:\:\\
0.41 \:\:& 0.70 \:\:& ... \:\:\\
\end{array}\right].
\end{array}\label{dt}
\end{equation}
In the table each column corresponds to a preparation $\mathcal{P}^{(j)}$. The probabilities should match those computed from quantum theory via the trace rule. Note that since there is a continuum of pure quantum states, our table has an infinite number of columns. I have kept the notation simple by pretending that the index parameterizing the preparations is discrete. It is clear that the table represents the two probabilistic outcomes of measurements (rows), which will of course depend on the preparation used (columns).

\subsection{Ontological factorizations of a data table}

We start with a composite data table $D$ corresponding to $m$ $d-$outcome measurements. The probabilistic outcomes are compiled in $m$ $d \times s$ matrices $D^{(x)}$, $x=1,2,...,m$. An \textit{ontological factorization} of data table $D$ \cite{HarriganRA07} consists of a factorization of each $D^{(x)}$ into the product of a $d \times \Omega$ measurement matrix $\mathcal{M}^{(x)}$ and an $\Omega \times s$ preparation matrix $\mathcal{P}$:
\begin{equation}
D^{(x)}=\mathcal{M}^{(x)}\mathcal{P},\;\;\;\;\forall x=1,2,\ldots
m.\label{MPequalsD}
\end{equation}
The matrices $\mathcal{M}^{(x)}$ and $\mathcal{P}$ must be non-negative and column-stochastic, that is, $0\le \mathcal{P}_{jk}\le 1$, and $\sum_k \mathcal{P}_{jk}=1$ for all $j$ (and similarly for $\mathcal{M}^{(x)}$).

Ontological factorizations are not unique and represent models, having the following interpretation. The $k$th column of $\mathcal{P}$ represents a classical probability distribution over $\Omega$ ontic states, which corresponds to the preparation procedure $\mathcal{P}^{(k)}$ of $D$. As for $\mathcal{M}^{(x)}$, $\mathcal{M}^{(x)}_{ij}$ is the probability of obtaining measurement outcome $i$, if the ontic state is $j$, when performing measurement $x$. Note that the complexity of the model (measured by the size of the matrices involved) increases with the number $\Omega$ of ontic states underlying it.

If all matrices $\mathcal{M}$ have entries which are either $0$ or $1$ the ontological factorization is called \textit{deterministic}, otherwise it is indeterministic. In \cite{HarriganRA07} Harrigan \textit{et al.} obtained a few general constructions of ontological factorizations for data tables. Let us call the number of measurements $m$, the number of preparations $s$, and the number of outcomes of each measurement $d$. Then in \cite{HarriganRA07} three OF constructions were presented: an indeterministic OF with $\Omega=s$ ontic states; a deterministic OF with $\Omega=d^m$ ontic states; and a more parsimonious (on $m$) deterministic OF with $\Omega=s(dm-1)$ ontic states. These OF's are general, that is, they represent valid models for arbitrary data tables.

OF's can be said to be \textit{economical} when they have fewer ontic states than the general models above. Economical models decrease the number of ontic states by exploring regularities in the data tables. In the case treated in the next sections, these regularities are linear inequalities satisfied by the probabilities arising from quantum measurements. 

The process of obtaining economical OF's was dubbed \textit{ontological compression}.  In what follows we will concern ourselves with a simple ontological compression scheme for data arising from a particular tomographically complete set of measurements on $d$-dimensional systems. We will start with Harrigan \textit{et al.}'s general model with $\Omega=d^m$ ontic states (which is valid even for $s=\infty$), and obtain economical OF's by carefully excluding some of the original ontic states. Let us start by analyzing the simplest such case, which consists of measurements on a 2-level system, i.e. a qubit.

\subsection{Example: one qubit}

In this section I illustrate the ideas reviewed above by presenting an ontological factorization (OF) for the data table (\ref{dt}), which refers to the set of all preparations for a single qubit, and measurements of the Pauli operators $X,Y,$ and $Z$. I chose to present this data table and its OF's here as an example to be followed for larger systems in the next section.

We could have chosen to study more general data tables for a qubit, including for example the continuum of possible projective measurements, as opposed to just the three Pauli operators. It is well known, however, that the probabilities associated with just three independent two-outcome observables are sufficient to calculate the probabilities associated with any of any other observable. Moreover, the eigenvectors of the Pauli operators form mutually unbiased bases, which are optimal for tomography.

Harrigan \textit{et al.}'s deterministic OF for this data table has $\Omega=8$ ontic states. The motivation for this particular OF is as follows. Experimenting with different preparations, one finds that there exist states yielding definite, repeatable results for each of the three measurements. These are, of course, the eigenstates of the $X, Y, Z$ observables. The most general model for three two-outcome experiments would allow for \textit{simultaneous} definite outcomes for these three observables, which of course is not allowed by quantum mechanics. Nevertheless, one can understand quantum mechanics as a constraint on these more general probabilistic models, an idea that has been around for a long time \cite{Holevo82, Hardy01, Barrett07, Mana04, BarnumBLW07}.

With this in mind, the set of ontic states I propose consists of all possible definite states for these three observables. Since each observable has two outcomes,  there are $2^3=8$ such ontic states. We can represent these ontic states by the columns of the $8 \times 8$ identity matrix; it is clear that arbitrary probability $n$-tuples can be written as  convex combinations of these 8 $8$-tuples. We need to assign a meaning to the position of the 1 in each ontic state, before we can present the matrices representing measurements. Thus, let $j$ be the position of the 1 in an ontic state ($j=1,2, ...,8$). In binary, $j-1$ is written as a three-bit string $b_3 b_2 b_1$, with bit $b_i$ indicating which of the two outcomes (labeled 0 or 1) should occur for a measurement of basis $i$. For example, $j=5$ corresponds to the ontic $n$-tuple $\vec{p}_5=(0,0,0,0,1,0,0,0)^T$. The binary representation of $j-1=4$ is $100$, indicating that this ontic state would yield definite outcomes of 0,0 and 1 for measurements of first, second and third bases, respectively.

The matrices representing these measurements were chosen according to the convention I used in numbering the ontic states. The three measurements are represented as follows:
\begin{eqnarray}
\mathcal{M}^{(1)}=\left(
\begin{array}{cccccccc}
1&0&1&0&1&0&1&0\\
0&1&0&1&0&1&0&1
\end{array}
\right),\\
\mathcal{M}^{(2)}=\left(
\begin{array}{cccccccc}
1&1&0&0&1&1&0&0\\
0&0&1&1&0&0&1&1
\end{array}
\right),\\
\mathcal{M}^{(3)}=\left(
\begin{array}{cccccccc}
1&1&1&1&0&0&0&0\\
0&0&0&0&1&1&1&1
\end{array}
\right).
\end{eqnarray}
As an illustration, suppose we want to measure $Z$ on a state (preparation) represented by the following one-column matrix $\mathcal{P}$:
\begin{equation}
\mathcal{P}=(\mathcal{P}_1,\mathcal{P}_2,\mathcal{P}_3,\mathcal{P}_4,\mathcal{P}_5,\mathcal{P}_6,\mathcal{P}_7,\mathcal{P}_8)^T,
\end{equation}
where $\mathcal{P}_i$ is the probability of preparation in ontic state $i$. The data table $D^{(3)}$ with the probabilities associated with outcomes 0 ($p_0$) and 1 ($p_1$) of $Z$ is given by the factorization
\begin{equation}
 D^{(3)}\equiv \left(
\begin{array}{c}
p_0\\
p_1
\end{array}
\right) =\mathcal{M}^{(3)}\mathcal{P}= \left(
\begin{array}{c}
\sum_{i=1}^4 \mathcal{P}_i\\
\sum_{i=5}^8 \mathcal{P}_i\
\end{array}
\right)
\end{equation}

Any data table $D^{(x)}$ corresponding to three two-outcome measurements can be given an OF using these eight ontic states - this is an example of one of the general constructions proposed in \cite{HarriganRA07}. For $m$ $d$-outcome measurements and $s$ preparations, the number of necessary ontic states is at most $\Omega=d^m$, but this number can be smaller, depending on the data table. We are interested in finding economical OF's for data arising from quantum measurements.

\begin{figure}[t]
\includegraphics[scale=0.25]{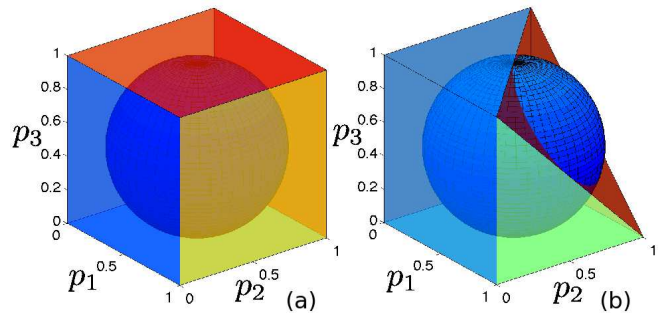}
\caption{ (Color online) (a) The cube $P_1$ represents general probability 3-tuples, which include the ball representing quantum states of a qubit. (b) If we eliminate the cube vertex $(1,1,1)$, the convex hull of the remaining 7 vertices is polyhedron $P_2$, which no longer contains all quantum states. All 8 vertices are required to account for all quantum states in this way.}
\label{figcube}
\end{figure}

Since the probabilities associated with the two outcomes of each observable must add up to 1, we can represent experimental outcomes by listing just one independent probability per observable, say the first one. For example, the first column in table (\ref{dt}) can be represented by the 3-tuple $\vec{p}=(0.93,0.73,0.59)$. In this 3-dimensional space of probabilities, general preparations are the convex hull of the eight ontic state vertices, which form a 0/1 cube in 3 dimensions which we will call polyhedron $P_1$. Quantum states are a proper subset of the cube, the ball of radius $r=1/2$ and with center $\vec{p}=(\frac{1}{2},\frac{1}{2},\frac{1}{2})$ (see Fig. \ref{figcube}-(a)).

To perform ontological compression, we need to find a smaller set of ontic states whose convex combinations still account for the data table. The simple ontological compression scheme I will consider consists of keeping only a proper subset of the original model's ontic states, and adapting the measurement matrices accordingly. For example, if we eliminate the eighth ontic state (cube vertex $\vec{p}=(1,1,1)$), the allowed preparations now can be convex combinations of only the 7 remaining ontic states. Geometrically, this is the convex hull of 7 vertices, the polyhedron $P_2$ represented in Fig. \ref{figcube}-(b). This deletion of one ontic state must be accompanied by a corresponding change in the measurement matrices $\mathcal{M}^{(1)}, \mathcal{M}^{(2)}$ and $\mathcal{M}^{(3)}$ -  the column associated with the deleted ontic state, the eighth, must also be deleted.

By comparing the new polyhedron $P_2$ with the ball of quantum states, we see that  there exist  quantum states $\vec{p} \notin P_2$. Hence, the proposed ontological compression scheme fails. Symmetry indicates the compression will also fail in case of removal of any other ontic state from our model. In this case we need the full set of eight ontic states if we are to represent all quantum states as convex combinations of a subset of the original set of ontic states (the cube vertices).

In the next section I will show that this failure of the simple ontological compression scheme I propose happens only for $d=2$. For $d \ge 3$ the scheme outlined above can greatly reduce the number of ontic states necessary to describe data tables arising from a tomographically complete set of measurements.

\section{Economical ontological factorizations}

In this section I show how to obtain ontological factorizations for data tables associated with all preparations of a $d$-dimensional quantum system, together with measurements onto a particular tomographically complete set of observables. I will start with the general, extravagant OF with $\Omega = d^m$ ontic states of Harrigan \textit{et al.} \cite{HarriganRA07}, and show how the number $\Omega$ of ontic states can be reduced, resulting in economical OF's.

The measurements we will consider are projectors onto a complete set of mutually unbiased bases (MUB) for a $d$-dimensional system. Such a set contains $d+1$ bases, each providing $d-1$ independent measurement probabilities. Such complete sets of MUB are known to exist for any Hilbert space dimension $d$ which is power-of-prime. In the case of prime-dimensional systems, there is a canonical construction of a complete set of MUB, due to Ivanovic \cite{Ivanovic81} - in the Appendix I list the four bases for $d=3$. We will use this canonical set, and restrict ourselves to discussing the case of prime $d$.

Our initial model is Harrigan \textit{et al.}'s deterministic OF \cite{HarriganRA07}, in which each measurement can yield any definite result. This OF must have as many ontic states as there are definite outcomes for the $d+1$ MUB measurements. Hence, the number of ontic states for this OF is $d^{d+1}$. We will denote the polytope formed by taking the convex hull of these ontic states $P_1$. The measurement matrices can be defined in a similar way as we did for a qubit in the last section, resulting in a deterministic OF.

In order to find OF's with fewer ontic states, we resort to some ideas from convex geometry. The $d^{(d+1)}$ ontic states can be chosen so that the $n$-tuple representing the $j$th ontic state has 0's in all positions, except the $j$th, which has 1. The set of allowed preparations are convex combinations of those ontic states, i.e. polytope $P_1$. If we remove ontic states from that initial set, the resulting OF will only account for preparations which lie on the convex hull of the remaining ontic states. This convex hull is a polytope, which we will call $P_2$. Our simple ontological compression scheme will involve finding subsets of the original set of ontic states which include all quantum states in their convex hull.

The identification of OF's with fewer ontic states is a way of approximating the convex set of quantum states by simpler convex bodies, the polytopes. The facets of the enclosing polytopes represent linear inequalities obeyed by any quantum state. These inequalities are the regularities we identify in the quantum data tables, enabling us to do the ontological compression. In this representation using probabilities associated with MUB, all pure quantum states lie on the surface of a high-dimensional ellipsoid \cite{Brukner99}.

Different approaches can be taken to perform this simple kind of ontological compression. In what follows I describe a general procedure that works for any $d \ge 3$, and exemplify the approach by tackling the qutrit case ($d=3$).

\subsection{A general compression scheme}\label{sec ddim}

Here I will point out one way of obtaining economical ontological factorizations (OF's) for $d$-dimensional systems, for prime $d$. First, we need to review how we can represent states using probabilities associated with mutually unbiased bases (MUB).

Not all components of the probability n-tuple $\vec{p}$ are equivalent: sets of $d-1$ components belong together, as they are independent probabilities associated with different outcomes of the same $d$-outcome observable. Let greek indices $\kappa, \mu$ etc denote the observable associated with a given set of probabilities. The representation of states in probability space would be with a $(d^2-1)$-tuple of probabilities $p_j^{(\kappa)}$:
\begin{equation}
\vec{p}=(p_1^{(1)}, p_2^{(1)}, ... , p_{(d-1)}^{(1)}, ... , p_{1}^{(d+1)}, ..., p_{(d-1)}^{(d+1)}), \label{eq pmub}
\end{equation}
where the super-indices indicate which of the MUB $\kappa$ the probabilities are associated with. Our initial OF consists of measurement matrices $\mathcal{M}$ that are trivial generalizations of the qubit case, and of the $d^{d+1}$ ontic states represented by probability vectors as above, where for each of the MUB $\kappa$ either all $p_j^{(\kappa)}=0$ (indicating that the omitted $p_{d}^{(\kappa)}=1$); or there is exactly one of the $p_j^{(\kappa)}=1$ for each $\kappa$, with the remaining $p_i^{(\kappa)}=0$. The convex hull of this set of $d^{d+1}$ ontic states is our initial polytope $P_1$. This choice of initial ontic states results in a polytope in which the probabilities $p_i^{(\kappa)}$ obey the MUB constraints $\sum_i p_i^{(\kappa)}=1$ (for all $\kappa$).

With a view to obtaining OF's which are as economical as possible, we can start with polytope $P_1$ and remove one randomly chosen vertex (a process that we can later iterate). The facets of the smaller resulting polytope $P_2$ can be found with a convex hull algorithm such as the one presented in \cite{BarberDH96}. To guarantee that all quantum states are contained in $P_2$, it is enough to consider each facet in turn, proving that no quantum state violates it. This can be done as follows.

For simplicity of notation, let us use a single sub-index $i$ to index the $(d^2-1)$ independent probabilities arising from MUB measurements that fully describe a $d$-dimensional quantum state. Each probability $p_i$ is obtained from the associated MUB projector $\hat{P}_i$ using the trace rule (eq. \ref{eqpi}).

Each facet is represented by a linear inequality on the MUB probabilities, and as such can be written as the expectation value of an operator built out of the associated projectors. The association is as follows:
\begin{equation}
\sum_i c_i p_i \ge f \Longleftrightarrow  \left\langle \sum_i c_i\hat{P}_i \right\rangle \ge f,
\end{equation} \label {ineqform}
where $f$ is a constant offset. Because all coefficients $c_i$ are real, the operator $\hat{C} \equiv \sum_i c_i \hat{P}_i$ is Hermitian. In this case, the minimax principle (\cite{Halmos74}, \S 90) guarantees that the minimum quantum expectation value of $\hat{C}$ will be given by its smallest eigenvalue $\lambda_{min}$. If $\lambda_{min} \ge f$ then no quantum state falls outside the facet.

\subsection{Example: $d=3$}\label{sec qutrit}

For $d=3$, I started with the 81-vertex polytope $P_1$ and tried to remove vertices randomly, checking whether all quantum states satisfy the inequalities representing the facets of the resulting polytope. This was done with the convex hull algorithm of \cite{BarberDH96} and simple eigenvalue calculations as described in the last section.

\begin{figure}[t]
\includegraphics[scale=0.5]{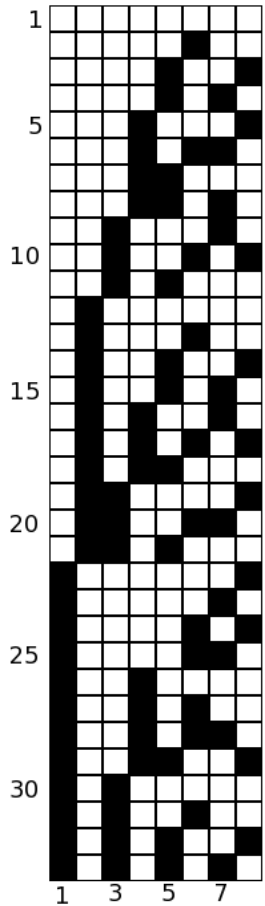}
\caption{The rows represent the 33 vertices of an 8-dimensional polytope that includes all qutrit states, where black and while code for 1 and 0, respectively.}
\label{fig_vertices}
\end{figure}

By iterating this procedure many times (backtracking when necessary), I obtained economical, deterministic OF's for a qutrit with as few as 33 ontic states. In Fig. \ref{fig_vertices} I list the vertices of one such OF. The probabilities are associated with the MUB projectors listed in the Appendix.

This 33-vertex polytope can be described also in terms of its facets. These are represented by 51 linear inequalities on the probabilities. Of these, 12 are the trivial inequalities $\sum_i p_i^{(\kappa)} \le 1$ (4 inequalities corresponding to $\kappa=1,2,3,4$) and $p_i^{\kappa}\ge0$ (8 inequalities corresponding to $\kappa=1,2,3,4$ and $i=1,2$ ). The remaining 39 inequalities are listed as the rows of Table \ref{table facets}. Each inequality is of the form $\sum_{i=1}^8 c_i p_i \ge f$. The last column of Table \ref{table facets}  presents the smallest eigenvalue $\lambda_{min}$ of operator $\hat{C} \equiv c_i \hat{P}_i$, where $\hat{P}_i$ is the projector associated with probability $p_i$. By the results of the last section, we see that no quantum state falls outside the polytope, and also that some of the facet inequalities are saturated by quantum states. At least some of the inequalities presented in Table \ref{table facets} are new, see \cite{PittengerR05} for a different set of similar linear inequalities, derived in finding the extrema of discrete Wigner functions for prime dimension $d$.

\begin{table}\caption{\label{table facets} Polytope facets} \begin{center}
\begin{tabular}{ | c | c | c | c | c | c | c | c | c | c |  }
\hline
$c_1$ & $c_2$ & $c_3$ &$c_4$ & $c_5$ & $c_6$&$c_7$&$c_8$ & $f$ & $\lambda_{min}$\\ \hline		
  -2&    -1&    -2&    -1&     1&    -1&    -1&    -2&    -5& -4.3028\\
  -2  &  -1 &   -2 &   -1&     2 &    1 &   -2 &   -1 &   -4 & -3.8608\\
    -2   & -1  &  -1 &   -2 &    2 &    1 &   -1 &   -2 &   -4 & -3.8608 \\
    -2    &-1   &  1 &   -1 &   -1 &    1 &   -1 &    1 &   -3 & -2.8608 \\
    -1     &0  &  -1 &   -1 &    1 &    0 &    0 &   -1 &   -2 & -2.0000 \\
    -1     &0  &  -1 &   -1 &    1 &    1 &   -1 &   -1 &   -2 & -2.0000 \\
    -1     &0   & -1 &    0 &    1 &    0&    -1 &   -1 &   -2 & -2.0000 \\
    -2     &1    &-3 &   -2 &    3 &    1 &   -2 &   -3 &   -5 & -4.8455 \\
    -1    &-2 &   -1&    -2 &   -2 &   -1&    -2&    -1&    -6 & -6.0000 \\
    -1    &-2 &   -1 &    1&    -1&    -2 &    1 &   -1 &   -4 & -4.0000 \\
    -1   & -2  &  -1 &    1&     1 &    2 &   -1 &    1 &   -2 & -2.0000 \\
    -1 &   -2  &   1 &   -1 &    1&     2&     1 &   -1&    -2 & -2.0000 \\
    -1    & 0    & 1 &    0&    -1 &    0 &    0 &    1 &   -1 & -1.0000 \\
    -1     &0    & 1 &    1 &   -1 &    0 &    1 &    1 &    0 & 0.0000 \\
    -1    & 1   & -1&    -2 &    1 &    2 &   -2 &   -1 &   -3 & -2.3028 \\
     0    &-1 &   -1 &   -1 &   -1 &    0 &   -1 &   -1 &   -3 & -3.0000 \\
     0    &-2  &  1 &   -1 &   -1 &   -2 &   -1 &    1 &   -4 & -3.7397 \\
     0   & -1 &    0 &   -1 &   -1 &   -1 &   -1 &    0 &   -3 & -2.9085 \\
     0   & -1&     0&    -1 &    0 &   -1 &    0 &    1 &   -2 & -1.9085 \\
     0  &  -1 &    1 &    0 &    0 &   -1 &   -1 &    0 &   -2 & -1.6180 \\
     0  &   1 &   -1 &    0 &    0 &   -1 &   -1 &    0 &   -2 & -1.9085 \\
     0    & 1 &    0 &   -1 &    0 &   -1 &    0 &   -1 &   -2 & -1.9085 \\
     0  &   1 &    0 &   -1  &   1  &   1 &   -1 &    0 &   -1 & -0.6180 \\
     1  &  -1 &   -2 &   -1 &   -1 &    1 &   -2 &   -1 &   -4 & -3.8608 \\
     1  &  -1 &   -1 &   -2 &   -1 &    1 &   -1 &   -2 &   -4 &-3.3028 \\
     1  &   0 &   -1 &   -1 &   -1 &    0 &    0 &   -1 &   -2 & -2.0000 \\
     1  &   0 &   -1 &   -1 &    1 &    0 &    1 &    1 &    0 & 0.0000 \\
     1   &  0 &   -1 &    0 &   -1  &   0 &   -1 &   -1 &   -2 & -2.0000 \\
     1   &  0 &    1 &    1 &    1 &    0 &   -1 &   -1 &    0 & 0.0000 \\
     1 &    1 &   -1 &    0 &   -1 &   -1 &   -1 &    0 &   -2 & -1.6180 \\
     1  &   1 &    0 &   -1 &   -1 &   -1 &    0 &   -1 &   -2 & -1.9085 \\
     1    & 1  &   0 &   -1 &    1 &    1  &   0 &    1 &    0 & 0.0915 \\
     1   &  1  &   1 &    0 &    1 &    1 &   -1 &    0 &    0 & 0.0915 \\
     1    & 2 &   -1 &    1 &   -1 &   -2 &   -1 &    1 &   -2 & -2.0000 \\
     1  &   2 &   -1 &    1 &    1  &   2  &   1  &  -1 &    0 & 0.0000 \\
     1   &  2  &   1 &   -1 &   -1 &   -2 &    1  &  -1 &   -2 & -2.0000 \\
     1   &  2  &   1 &   -1 &    2 &    2 &   -1  &   1 &    0 & 0.2603 \\
     2  &  -1 &   -3 &   -2 &   -3 &   -1 &   -2 &   -3 &   -7 & -6.8455 \\
     2   &  1  &  -2 &   -1 &   -1 &    1 &   -1 &   -2 &   -3 &-2.8608\\
\hline  
\end{tabular}
\end{center}
\end{table}

I have also checked that this 33-vertex polytope is minimal, in the sense that removal of any of its 33 vertices results in a polytope that no longer includes all quantum states in its interior, i.e. its ontic states no longer account for all quantum states as convex combinations. There are many other minimal polytopes (with varying number of vertices) representing economical OF's for a qutrit. A limited random search failed to find any such OF's with less than 33 ontic states.

This computational approach to finding economical OF's can complement other approaches in the search for ontological models with as few ontic states as possible. For example, one may start with the construction of ontological models recently delineated in \cite{DakicSPB08} and apply the random pruning described here to decrease the number of ontic states necessary in an ontological model for tomographically complete sets of quantum measurements.

\section{Conclusion}

I have reviewed the ontological model formalism proposed in \cite{HarriganRA07}, and described a procedure that results in economical models for a particular set of tomographically complete measurements on general states of a $d$-dimensional quantum system. To illustrate the procedure, I brought down the number of necessary ontic states from $81$ to just $33$ in the case of a qutrit ($d=3$). In the process, I found some new linear inequalities for probabilities associated to mutually unbiased bases for a qutrit.

\begin{acknowledgments}
I would like to thank Ingemar Bengtsson and Karol \.{Z}yckowski for helpful discussions, and acknowledge support from Brazilian funding agencies FAPERJ and CNPq.
\end{acknowledgments}

\appendix*

\section{}
In this Appendix I list the four mutually unbiased bases (MUB) for a qutrit that we used in section \ref{sec qutrit}. The probabilities $p_i^{(\kappa)}$ of measuring density matrix $\rho$ onto the $i$th projector of basis $\kappa$ are defined as:
\begin{equation}
p_i^{(\kappa)}\equiv Tr\left(\rho \left| v_i^{(\kappa)}\right\rangle \left\langle v_i^{(\kappa)} \right| \right).\label{eqpi}
\end{equation}

Using $\omega \equiv \exp{(2\pi i/3)}$, the MUB vectors can be written as: 
\begin{eqnarray}
v_1^{(1)}&=&(0,1,0)\\
v_2^{(1)}&=&(0,0,1)\\
v_3^{(1)}&=&(1,0,0)\\
v_1^{(2)}&=&1/\sqrt{3}(1,1,1)\\
v_2^{(2)}&=&1/\sqrt{3} \left(\omega,1,\omega^2 \right)\\
v_3^{(2)}&=&1/\sqrt{3} \left(\omega^2,1,\omega \right)\\
v_1^{(3)}&=&1/\sqrt{3} \left(\omega,1,1\right)\\
v_2^{(3)}&=&1/\sqrt{3} \left(1,\omega,1\right)\\
v_3^{(3)}&=&1/\sqrt{3} \left(1,1,\omega \right)\\
v_1^{(4)}&=&1/\sqrt{3} \left(\omega^2,1,1\right)\\
v_2^{(4)}&=&1/\sqrt{3} \left(1,\omega^2,1\right)\\
v_3^{(4)}&=&1/\sqrt{3} \left(1,1,\omega^2 \right)
\end{eqnarray}

This is the $d=3$ case of the canonical construction of complete sets of mutually unbiased bases for prime-dimensional quantum systems due to Ivanovic \cite{Ivanovic81}.

\bibliographystyle{unsrt}

\end{document}